# ELEMENTARY COMPUTATION AND VON NEUMANN ALGEBRAS

MARCO PEDICINI AND MARIO PIAZZA

Abstract. In this paper, we show how a construction of an implicit complexity model can be implemented using concepts coming from the core of von Neumann algebras. Namely, our aim is to gain an understanding of classical computation in terms of the *hyperfinite* $\mathrm{II}_1$ *factor*, starting from the class of Kalmar recursive functions. More methodologically, we address the problem of finding the right perspective from which to view the new relation between computation and combinatorial aspects in operator algebras. The rich structure of discrete invariants may provide a mathematical setting able to shed light on some basic combinatorial phenomena that are at the basis of our understanding of complexity.

## 1. Introduction

The theory of operator algebras, one of the fundamental achievements of twentieth century mathematics, is perhaps the most general example of a mathematical theory within which *discrete* structures do manipulate *continuous* ones. Moreover, discrete structures in the theory of operator algebras (e.g., discrete symmetry groups) are very special, as they emerge ubiquitously in culturally distant areas of mathematics. Since logic and computation are by nature discrete, the important question arises whether discrete properties as developed in operator algebras may provide a conceptual tool for tracking what is logical inside mathematics and for explaining complexity levels of computation as well. Within such broad perspective, it has been proposed to reshape the semantics of computation called *geometry of interaction* (GoI) looking at logic as a bundle of fundamental computational processes to be recaptured in the operator algebras setting [Gir06a].

The aim of this paper is to show how the theory of von Neumann algebras can be used in a transparent way to obtain an implicit complexity model of classical Turing machines (TM). More methodologically, we address the problem of finding the right perspective from which to view the new relation between computation and combinatorial aspects in operator algebras.

It is well-known that the foundation of operator algebra in the work of von Neumann and Murray is intertwined to the interpretation of quantum mechanics. There are many definitions of von Neumann algebra. The most common one considers a von Neumann algebra as a set of operators acting on some given Hilbert space. More formally, a von Neumann algebra is a unital selfadjoint subalgebra of the algebra of all bounded linear operators on a Hilbert space which is closed under the topology of pointwise convergence. In this paper, we explore the rich internal structure of von Neumann algebras through factors, i.e. von Neumann algebras whose center consists of scalar multiples of the identity. They can be viewed as the elementary constituents from which all the von Neumann algebras are built: every von Neumann algebra is a direct integral (a generalization of direct sum) of factors [MvN36, MvN37, MvN43, vN49].

In particular, we focus our attention on the unique von Neumann algebra which admits finite dimensional approximations and it is a factor $\text{II}_1$: the so called hyperfinite $\text{II}_1$ factor $\mathcal{R}$. By "finite dimensional approximation", we mean that each finite subset of the algebra can be approximated, to arbitrary precision, by an element from a finite-dimensional subalgebra. Traditional GoI [Gir89, Gir90, Gir95, Gir06b] is built in a $C^*$-algebra: proofs correspond to bounded operators of the infinite dimension Hilbert space, and the execution formula corresponds to the power series of the operator itself. Even if a von Neumann algebra is automatically a $C^*$-algebra, it is conceptually misleading to conceive von Neumann algebras merely as instances of $C^*$-algebras. In fact, it is an old idea that $C^*$-algebras correspond to noncommutative *topological spaces*, while von Neumann algebra corresponds to noncommutative *measure spaces* [Bla06, p. 221]. Moreover, what is crucial to the approach to logical complexity via von Neumann algebras is the quest for mathematical *universality* and *invariance*. In this light, it is of great significance to work on an universal object such as the *unique* $\mathcal{R}$ (two any hyperfinite $\text{II}_1$ factors are isomorphic), which is contained in any $\text{II}_1$ factor.

The paper is organized as follows. Section 2 contains various definitions, important facts and results on von Neumann algebras which constitute the mathematical framework of the paper. The setting is a classical one, where 'classical' essentially means 'commutative'. So, we consider an *abelian* group which has a simpler structure and which may be viewed, from a methodological point of view, as a main tool to construct an embedding of classical computational devices in the setting of von Neumann algebras. Moreover, the case of *von Neumann group algebras* built upon a commutative group is quite special and it was studied at the very beginning of the development of the theory of operator algebras, also thanks to the conceptual pervasiveness of Pontryagin duality. Here these notions are employed in concert to show that the group von Neumann algebra of $G$, denoted by $\mathcal{N}(G)$, is a *maximal self-adjoint abelian subalgebra* in $\mathcal{R}$ (MASA), another universal property which our model of computation utilizes. In Section 3, we present the construction of an ascending sequence $(G_i)$ of finite cyclic groups such that the cardinality of $G_i$ doubles at every step, and whose infinite union is a discrete group $G$. In Section 4 we present the setting for our model of elementary computation. An elementary function can be computed by a TM with an elementary execution time and with a similar elementary bound on the space: the bound on the space implies a similar (elementary) bound on the number of distinct configurations. We tackle the task of encoding a TM into operators acting on von Neumann group algebra $\mathcal{N}(G)$ of the group $G$. To this aim we introduce the auxiliary notions of *computational trees* and *branches.* Finally, we propose in passing an estimation of the number of the possible initial configurations for a TM corresponding to an $n$-ary input vector of integers with a unary representation of length $L$. In the last Section 5, we embed the configurations of a TM in the Hilbert space $\ell^2(G)$ of the square summable formal series indexed by elements of $G$ with complex coefficients. Because the number of configurations in a finite part of the tape is bounded, clearly it is possible to give a mapping which sends the set of configurations to a finite dimensional subspace. Hence we are able to model the transition function of the machine by means of an endomorphism acting on the subspace, so as to prove the claim that any elementary function is representable in a subspace of $\mathcal{R}$.

## 2. Preliminaries and basic results

In this section we set up the notation, recall some basic results, and state some important results which constitute the mathematical core of the paper.

The theory of finite (and hyperfinite) factors relies on the notion of the (infinite) *separable Hilbert space*, that is a Hilbert space with a denumerable orthonormal basis. This means that if the basis is $G$, i.e. a denumerable discrete group (by construction), then the resulting space is automatically separable.

Let $G$ be a (topological discrete) group and let $\mathbb{C}[[G]]$ denote the set of all functions from $G$ to $\mathbb{C}$ expressed as formal sums (that is, a function $a : G \to \mathbb{C}$, $g \mapsto a(g)$, is written as $\sum_{g\in G} a(g)g$). Let $\mathbb{C}G$ denote the complex group ring of formal sums with finite support. Then, for each $a \in \mathbb{C}[[G]]$, we define

$$||a|| := (\sum_{g\in G} |a(g)|^2)^{1/2} \in [0, \infty],$$

and $tr(a) := a(1) \in C$.

Then, let us define

$$\ell^2(G) := \{a \in \mathbb{C}[[G]] : ||a|| < \infty\}.$$

Let $\mathbb{C}G$ denote the complex group ring of formal sums with finite support. Then, we view $\mathbb{C} \subset \mathbb{C}G \subset \ell^2(G) \subset \mathbb{C}[[G]]$. There is a well-defined external multiplication map

$$\ell^2(G) \times \ell^2(G) \to \mathbb{C}[[G]], \quad (a, b) \mapsto a \cdot b,$$

where, for each $g \in G$, $(a \cdot b)(g) := \sum_{h\in G} a(h)b(h^{-1}g)$; this series converges in $\mathbb{C}$, and, moreover, $|(a \cdot b)(g)| \leq ||a|| \, ||b||$, by the Cauchy-Schwarz inequality. The external multiplication extends the multiplication of $\mathbb{C}G$.

Then, $\ell^2(G)$ is a separable Hilbert space with the scalar product defined by

$$\langle \sum_{g\in G} \lambda_g g, \sum_{g\in G} \mu_g g\rangle = \sum_{g\in G} \lambda_g \overline{\mu}_g$$

where $\overline{\mu}$ denotes the complex conjugate of $\mu$.

Let us denote the space of the bounded operators in the Hilbert space $\ell^2(G)$ by $\mathcal{B}(\ell^2(G))$. And let $\mathcal{U}(\ell^2(G)) \subset \mathcal{B}(\ell^2(G))$ be the subset of unitary operators (operators such that $uu^* = u^*u = 1$) from $\ell^2(G)$ to $\ell^2(G)$.

We define the left regular representation $\lambda'$ of $G$ as the function $\lambda' : G \to \mathcal{U}(\ell^2(G))$ such that the image of an element of the group $G$ is an unitary operator $\lambda'(g) : \ell^2(G) \to \ell^2(G)$ so that

$$\lambda'(g) \sum a_h h := \sum a_h gh.$$

Finally, we denote with $\lambda : \mathbb{C}G \to \mathcal{B}(\ell^2(G))$ the extension of the left regular representation $\lambda'$ of $G$ in

$$\lambda : \mathbb{C}G \to \mathcal{B}(\ell^2(G))$$

where

$$\lambda(\sum_{g\in G} a(g)g) := \sum_{g\in G} a(g)\lambda'(g).$$

We recall that the *von Neumann group algebra of* $G$, denoted by $\mathcal{N}(G)$, is the closure of $\lambda(\mathbb{C}G)$ in the strong operator topology on $\mathcal{B}(\ell^2(G))$.

Such object can be described as the algebra of (right) $G$-equivariant bounded operators from $\ell^2(G)$ to $\ell^2(G)$:

$$\mathcal{N}(G) = \{\alpha : \ell^2(G) \to \ell^2(G) \,|\, ||\alpha|| < +\infty, \text{ for any } h \in \ell^2(G) \text{ and } g \in G, \, \alpha(h)g = \alpha(hg)\}.$$

Now we are ready to recall the basic facts on factors and in particular the definition of hyperfinite $\mathrm{II}_1$ factor $\mathcal{R}$. Such definition is based upon the notion of projection and the typical "abundance of projections" in von Neumann algebras.

Let $\mathcal{H}$ be a Hilbert space and $\mathcal{B}(\mathcal{H})$ the bounded operators on $\mathcal{H}$. The *commutant* of the von Neumann algebra $M$ contained in $\mathcal{B}(\mathcal{H})$ is

$$M' = \{x' \in \mathcal{B}(\mathcal{H}) \text{ s.t. } x'x = xx' \text{ for all } x \in M\}.$$

**Definition 1.** A von Neumann algebra $M$ is called a *factor* if it has trivial center, i.e. $Z(M) = M \cap M' = \mathbb{C}I$.

The concept of projection plays a fundamental role in the study (and the classification) of factors. A *projection* is any element $p \in M$ such that $p = p^* = p^2$. A *sub-projection* of a projection $e$ is any projection $f$ such that $ef = f$. It follows that if $M$ is a factor, then any two projections are comparable in the sense that one is equivalent to a sub-projection of the other. The equivalence $e \simeq f$ is given in the sense of Murray-von Neumann, namely it holds if (and only if) there exists a partial isometry $u$ such that $e = u^*u$ and $f = uu^*$.

A projection is *finite* when it is not equivalent to any of its proper sub-projections. Moreover, a factor $M$ is said to be of type II if $M$ contains non-null finite projections and there exists no non-null minimal projection in $M$. A type II factor is said to be of type $\text{II}_1$ if it does not contain any non unitary isometry (i.e. it is a finite factor). Now, within the class of all $\text{II}_1$ factors, there is one, the so-called *hyperfinite* $\text{II}_1$ *factor* $\mathcal{R}$, uniquely determined up to isomorphism by the property of being approximately finite dimensional, [MvN43]. A von Neumann algebra $M$ is said to be approximately finite dimensional if there exists an increasing sequence

$$A_1 \subset A_2 \subset \cdots \subset A_n \subset A_{n+1} \subset \ldots$$

of finite dimensional $*$-subalgebras, whose union $\bigcup_{i=0}^{\infty} A_i$ is $\sigma$-weakly dense in $M$.

Whenever the discrete group $G$ is obtained by an approximation procedure, i.e., by an increasing sequence of finite subgroups

$$G_0 \subset G_1 \subset G_2 \subset \cdots \subset G_n \subset \ldots$$

such that $G = \bigcup_{i=0}^{\infty} G_i$, we say $G$ is *locally finite.*

Then, the von Neumann group algebra $\mathcal{N}(G)$ is approximately finite dimensional as stated in the following lemma:

**Lemma 2.1.** $\bigcup_{i=0}^{\infty} \mathcal{N}(G_i)$ *is weakly dense in the von Neumann group algebra of* $G = \bigcup_{i=0}^{\infty} G_i$.

*Proof.* We sketch a proof of this well established claim in the study of factors. If we consider a subgroup $H$ of $G$, by the left regular representation $\lambda'_G$ of $G$ in $\mathcal{N}(G)$, we may embed any element of $H$ in $\mathcal{N}(G)$. By considering the double-commutant of $\lambda'_G(H)$, we obtain a von Neumann group algebra which is sub-algebra of $\mathcal{N}(G)$.

When we consider $\bigcup_i \mathcal{N}(G_i)$ since we have an isomorphism of algebras between $\mathcal{N}(G_i)$ and the subalgebras $(\lambda'_{G_i})''$ of $\mathcal{N}(G)$, we infer that $\bigcup_i \mathcal{N}(G_i)$ is isomorphic to $L = \bigcup_i (\lambda'_{G_i})'' \subset \mathcal{N}(G)$, [KR97]. Moreover $\mathcal{N}(G)$ is generated by $\lambda'_G(G)$ and for any $g \in G$, $\lambda'(g) \in L$, thus $L$ contains all the generators of $\mathcal{N}(G)$. Therefore, $\mathcal{N}(G)$ is a subalgebra of $L$: this concludes the proof since $L \subset \mathcal{N}(G) \subset L''$.

□

A traditional way to obtain an isomorphic copy of $\mathcal{R}$ is based on the notion of ICC group, where the acronym "ICC" stands for "infinite conjugacy class":

**Definition 2.** A discrete group $G$ is an *ICC group* if every non-trivial conjugacy class $C(h) = \{ghg^{-1} | g \in G\}$, for $h \neq 1_G$, is infinite.

Such property is sufficient to obtain a $\mathrm{II}_1$ factor, but insufficient to get a *hyperfinite* one (as it is the case when we consider free groups generated by $n > 1$ generators which are ICC groups but not locally finite):

**Proposition 2.2.** *If $G$ is an ICC group, then $\mathcal{N}(G)$ is a $\mathrm{II}_1$ factor.*

**Proposition 2.3.** *If $G$ is a locally finite ICC group, then $\mathcal{N}(G)$ is a hyperfinite $\mathrm{II}_1$ factor.*

The classical example of a locally finite ICC group is given by the group $S_X$ of permutations $\pi : X \to X$ of a countable set $X$ such that $\pi(x) = x$ for all $x \in X$ but a finite subset of $X$. By Proposition 2.3, it follows readily that $\mathcal{N}(S_X)$ is a hyperfinite $\mathrm{II}_1$ factor, see [Bla06].

With this preparation behind us, we are now ready to state the main theorem of the first part of the paper. The theorem can be achieved by suitably composing well-established facts and arguments in the theory of factors.

**Theorem 2.4.** *For any infinite denumerable discrete abelian group $G$, any von Neumann group algebra $\mathcal{N}(G)$ of $G$ is isomorphic to every maximal abelian subalgebra of the hyperfinite $\mathrm{II}_1$ factor $\mathcal{R}$.*

*Proof.* The proof splits into three parts corresponding to the following three sections: in Section 2.1, we observe that, in the commutative case, any von Neumann algebra can be classified up to isomorphism with respect to its minimal projections. In particular, up to isomorphism there exists a unique von Neumann algebra with a denumerable set of generators and without minimal projections (Corollary 2.6). Note also that isomorphisms of algebras preserve projections and the order between projections.

Our next aim, in Section 2.2, is to prove that $\mathcal{N}(G)$ has no minimal projection, Lemma 2.8. Furthermore, we can prove also that any MASA of $\mathcal{R}$ has no minimal projection, Lemma 2.9. Thus, in Section 2.3 we are in position to conclude that $\mathcal{N}(G)$ is (isomorphic) to this sole MASA of $\mathcal{R}$. □

*Remark* 1. It is important to point out that the accomplished proof of this fact is irrespective of a particular choice of the von Neumann algebra. This means that in order to embed the von Neumann group algebra $\mathcal{N}(G)$ it is sufficient to consider any von Neumann algebra with separable predual and without minimal projection. So apart from the case at issue, the hyperfinite $\mathrm{II}_1$ factor $\mathcal{R}$, the same argument applies to any other $\mathrm{II}_1$ factor and to factors of type $\mathrm{III}_1$ as well.

2.1. **Isomorphisms of abelian von Neumann algebras.** Commutativity of the algebra provides a way for the classification based on the structure of its projections. The classification theorem can be applied to any abelian von Neumann algebra, for there always exists an isomorphism with a maximal abelian von Neumann algebra. Then, any maximal abelian von Neumann algebra is characterized up to unital isomorphisms, [KR97, Thm. 9.4.1].

The symbol $\mathcal{H}_c$ stands for the separable Hilbert space $L^2 := L^2([0,1], \mathcal{S}, m)$, where $\mathcal{S}$ is the $\sigma$-algebra of Borel set of the interval $[0,1]$ and $m$ is the Lebesgue measure. With $f \in L^\infty$, $M_f$ is the operator of multiplication by $f$ on $L^2$.

**Definition 3.** Let us denote by $\mathcal{A}_c$ the set $\{M_f : f \in L^\infty\}$, which is a maximal abelian von Neumann algebra acting on $\mathcal{H}_c$ and without minimal projections.

The case of algebras having minimal projections is determined by the cardinality of these projections:

**Definition 4.** Let be $j$ any positive integer $j$ or $\omega$ itself, and let $\mathcal{S}_j$ be a set with $j$ elements, and $\mathcal{H}_j$ be the $j$-dimensional Hilbert space $\ell^2(\mathcal{S}_j)$. For each bounded function $f$ on $\mathcal{S}_j$, the multiplication by $f$ is a bounded linear operator $M_f$ on $\mathcal{H}_j$ and

$$\mathcal{A}_j := \{M_f : f \in \ell^\infty(\mathcal{S}_j)\}$$

is a maximal abelian von Neumann algebra acting on $\mathcal{H}_j$.

It can be proved that $\mathcal{A}_j$ has exactly $j$ minimal projections which generate $\mathcal{A}_j$. It follows that each of $\mathcal{A}_c$, $\mathcal{A}_j$, $\mathcal{A}_c \oplus \mathcal{A}_j$ is a maximal abelian von Neumann algebra acting on a separable Hilbert space. Moreover, no two of them are $*$-isomorphic since $\mathcal{A}_c$ has no minimal projection, $\mathcal{A}_j$ has exactly $j$ projections which generate it, and $\mathcal{A}_c \oplus \mathcal{A}_j$ has exactly $j$ minimal projections which do not generate it. So the remarkable result we are interested in is the following [KR97, Thm. 9.4.1].:

**Theorem 2.5.** *Each maximal abelian von Neumann algebra acting on a separable Hilbert space is unitarily equivalent to exactly one of the algebras:*
*a)* $\mathcal{A}_c$*;*
*b)* $\mathcal{A}_j$ *for all* $j \in \omega$ *or* $j = \omega$*;*
*c)* $\mathcal{A}_c \oplus \mathcal{A}_j$ *for all* $j \in \omega$ *or* $j = \omega$*.*

In particular, as we said above, we are concerned with the case of abelian algebras without minimal projections (letter $a$) of the above theorem), for which the following corollary holds:

**Corollary 2.6.** *If an abelian von Neumann algebra acting on a separable Hilbert space has no minimal projection, then it is isomorphic to* $\mathcal{A}_c$*.*

In fact, any abelian von Neumann algebra acting on a separable Hilbert space is isomorphic (not necessarily via a unitary isomorphism) to a maximal abelian von Neumann algebra. Thus by applying Theorem 2.5 we obtain the result.

In the next section, we show that the von Neumann group algebra $\mathcal{N}(G)$ where $G$ is an abelian infinite denumerable discrete group, offers a case in point. Namely, it has no minimal projection (by construction, it acts on a separable Hilbert space, i.e., $\ell^2(G)$). To this aim, we need to introduce another piece of our mathematical apparatus: a (small) part of the theory of group representations.

2.2. **Locally compact groups and Haar measure.** Broadly speaking, a duality theory shows the possibility to reconstruct a group from its representations. A representation is a description of the group in terms of linear transformations of vector spaces; more specifically, we use the representation in order to have the group operation represented by matrix multiplication.

We recall the fact that any *abelian* infinite group $G$ endowed with the discrete topology is a *locally compact group.* It is then possible to consider the set of all continuous homomorphism from $G$ to the circle group $\mathbb{T}$: these homomorphisms are called the *characters* of $G$. The set of characters of $G$, denoted by $\hat{G}$, with multiplication defined pointwise, is an abelian group called the (Pontryagin) *dual group.*

Note that under the discrete topology all singletons of $G$ are closed (and also open) sets so that any subset of $G$ is a Borel set. A measure on $G$ which is invariant by translation on Borel sets:

$$\mu(E) = \mu(g \cdot E) \qquad \text{for any Borel set } E \text{ and any } g \in G,$$

is achieved by giving the same value to any singleton set $\mu(\{g\}) = \alpha$ and it is unique up to the constant factor $\alpha$. This measure is called the *Haar measure* on $G$.

Thanks to the discrete measure, $L^2(G, \mu) = \ell^2(G)$. Moreover, the dual group $\hat{G}$ is endowed with pointwise convergence topology and the Haar measure $\hat{\mu}$. We also consider the well-known fact that if $\hat{G}$ is a compact set, then the measure of $\hat{G}$ is *finite*.

**Definition 5.** An *atom* $A$ in a measure space is a measurable set such that any measurable subset $B \subset A$ has either measure zero or the same measure as $A$.

Any atom $A$ is a singleton $x$ in $\hat{G}$. To check this fact, it is enough to observe that $\hat{G}$, as the dual of a discrete denumerable group $G$, is a separable compact group. Consequently, it has metric (and it is complete with respect to any metric compatible with its topology). Then, we may apply the following lemma [Roy88, pag. 408]:

**Lemma 2.7.** *Let $\mu$ be a Borel measure on a separable metric space $X$ and $A$ an atom for $\mu$. Then, there is a point $x \in A$ such that $\mu(A/x) = 0$.*

In fact, any atom is a singleton, thus any singleton is measurable and, by invariance of the measure, there exists infinite many measurable sets of non-zero measure. But in this way we get a contradiction, for, by $\sigma$-additivity, it results that $\hat{\mu}(\hat{G})$ is infinite.

Since any projection $p$ in the algebra $L^\infty(\hat{G}, \hat{\mu})$ is (equivalent to) the characteristic function of a measurable set, the order relation between projections $p < q$ corresponds to the inclusion of the sets of the corresponding characteristic functions. More precisely, let $p \equiv_\mu \chi_A$ and $q \equiv_\mu \chi_B$ (where $f \equiv_\mu g$ means $f(x) = g(x)$ for all $x \in \ell^2(\hat{G})$ up to a null measure set). Then, since $p < q$ is defined as $pq = q$ then $\chi_A \chi_B \equiv_\mu \chi_B$ but $\chi_A \chi_B \equiv_\mu \chi_{A \cap B}$; in this way, if and only if $A \subset B$ we have $\chi_{A \cap B} \equiv_\mu \chi_B$. Therefore, any atom corresponds to the notion of minimal projection in the sense that its characteristic function is a projection, and it is minimal because any subset of the atom has null measure. Hence there is no (non trivial) sub-projection.

This proves the following lemma:

**Lemma 2.8.** *For every infinite discrete abelian group $G$ the abelian von Neumann algebra $\mathcal{N}(G)$ has no minimal projection.*

We are now in position to reach the final conclusion.

2.3. **MASA of $\mathcal{R}$.** In fact, when we consider a MASA $M$ of $\mathcal{R}$, we have that $M$ acts on a separable Hilbert space insofar as it is a subalgebra of $\mathcal{R}$. Since the algebra $M$ is abelian, by applying the classification theorem, in particular Corollary 2.6, it follows that it is isomorphic to $\mathcal{A}_c$.

Once again, the main point here is to prove that this algebra has no minimal projection since the rest (the separability) is encapsulated in the very definition.

**Lemma 2.9.** *If $M$ is a MASA in $\mathcal{R}$, then $M$ has no minimal projection.*

*Proof.* In order to get a contradiction, let us suppose that $p \in M$ is minimal. Then, by the fact that $\mathcal{R}$ is of type II, $p$ cannot be minimal in $\mathcal{R}$, and so there exists a non null projection $q$ such that $qp = q$. Evidently $q \notin M$, thanks to the minimality hypothesis. On the other hand, $q$ commutes with $M$, contradicting the maximality of $M$. □

From the previous lemma, we infer that $M$ is isomorphic to $\mathcal{A}_c$ and from transitivity we get the following:

**Proposition 2.10.** *If $A_1$ and $A_2$ are MASA is $\mathcal{R}$, then $A_1$ is isomorphic to $A_2$.*

*Proof.* Thus by applying twice Corollary 2.6 we have obtained that $\mathcal{N}(G)$ is isomorphic to $\mathcal{A}_c$ and in turn that it is isomorphic to (any) MASA of $\mathcal{R}$. It follows that $\mathcal{N}(G)$ can be embedded in a MASA of $\mathcal{R}$ up to isomorphisms. □

By Zorn's Lemma, it is possible to build a maximal abelian subalgebra of $\mathcal{R}$, and this concludes the proof of the Theorem 2.4.

## 3. A construction of an AFD von Neumann algebra

In this section, we present an instance of the construction of the $G_i$ such that the relative dimension of two successive groups is 2, i.e., they have Lagrange index $[G_i : G_{i+1}] = 2$.

As a consequence of the previous section we confine ourselves to the case of a locally finite abelian group $G$, for we have by Theorem 2.4, that there exists an embedding $J : \mathcal{N}(G) \to \mathcal{R}$. Moreover, this embedding can be restricted on the subalgebras $\mathcal{N}(G_i)$, so that $J_i : \mathcal{N}(G_i) \to \mathcal{R}$.

We provide the explicit construction of an abelian locally finite group $G$. More specifically, we define $G_{i+1} = G_i + \mu_i G_i$ where $\mu_i$ is a new generator. The sequence starts at $G_0$, the trivial group containing the sole identity, here denoted by 0, $G_1 := C_2$ is the cyclic group of 2 elements, and in general $G_i := C_{2^i}$. Cyclic groups are abelian with one group generator. We give a construction of the multiplication table of $G_i$ in such a way that its group generator coincides with the element $\mu_i = 2^i$. In fact, this construction is recursive and uses an auxiliary permutation matrix $T(i)$, which generalizes the twist and gives an isomorphic presentation of $C_{2^i}$:

$$T(i) := \begin{cases} \begin{pmatrix} 1 \end{pmatrix} & \text{if } i = 0 \\ \begin{pmatrix} 0 & I_{2^{i-1}} \\ T(i-1) & 0 \end{pmatrix} & \text{otherwise.} \end{cases} \tag{3.1}$$

In a similar way we give the construction of the multiplication table of $G_i$:

$$G_i := \begin{cases} \begin{pmatrix} 0 \end{pmatrix} & \text{if } i = 0, \\ \begin{pmatrix} G_{i-1} & 2^{i-1} + G_{i-1} \\ 2^{i-1} + G_{i-1} & T(i-1).G_{i-1} \end{pmatrix} & \text{otherwise.} \end{cases} \tag{3.2}$$

The main feature of this construction is that any $G_i$ is subgroup of $G_{i+1}$ and it appears directly in its multiplication table as in the top-left corner. Another noticeable property is that the generator of the group is the element $2^i$.

In Figure 1, we have depicted with grey tones the multiplication tables of several $G_i$'s, in order to graphically point out the structure of these groups. The first groups are:

$$G_1 = \begin{pmatrix} 0 & 1 \\ \underline{1} & \underline{0} \end{pmatrix} \qquad G_2 = \begin{pmatrix} 0 & 1 & 2 & 3 \\ 1 & 0 & 3 & 2 \\ \underline{2} & \underline{3} & \underline{1} & \underline{0} \\ 3 & 2 & 0 & 1 \end{pmatrix} \qquad G_3 = \begin{pmatrix} 0 & 1 & 2 & 3 & 4 & 5 & 6 & 7 \\ 1 & 0 & 3 & 2 & 5 & 4 & 7 & 6 \\ 2 & 3 & 1 & 0 & 6 & 7 & 5 & 4 \\ 3 & 2 & 0 & 1 & 7 & 6 & 4 & 5 \\ \underline{4} & \underline{5} & \underline{6} & \underline{7} & \underline{2} & \underline{3} & \underline{1} & \underline{0} \\ 5 & 4 & 7 & 6 & 3 & 2 & 0 & 1 \\ 6 & 7 & 5 & 4 & 1 & 0 & 3 & 2 \\ 7 & 6 & 4 & 5 & 0 & 1 & 2 & 3 \end{pmatrix}$$

The underlined element of the table is $\mu_i$ and let us consider $\mu_3 = 2^2 = 4$ and its row in the multiplication table. With the aid of this table one may compute the different orbits for the iterated action of $\mu_i$ over elements of $G_i$,

$$O_i(x) := (x, \mu_i \cdot x, \mu_i \cdot \mu_i \cdot x, \dots, \mu_i^{2^i-1} \cdot x), \qquad \text{for all } x \in G_i$$

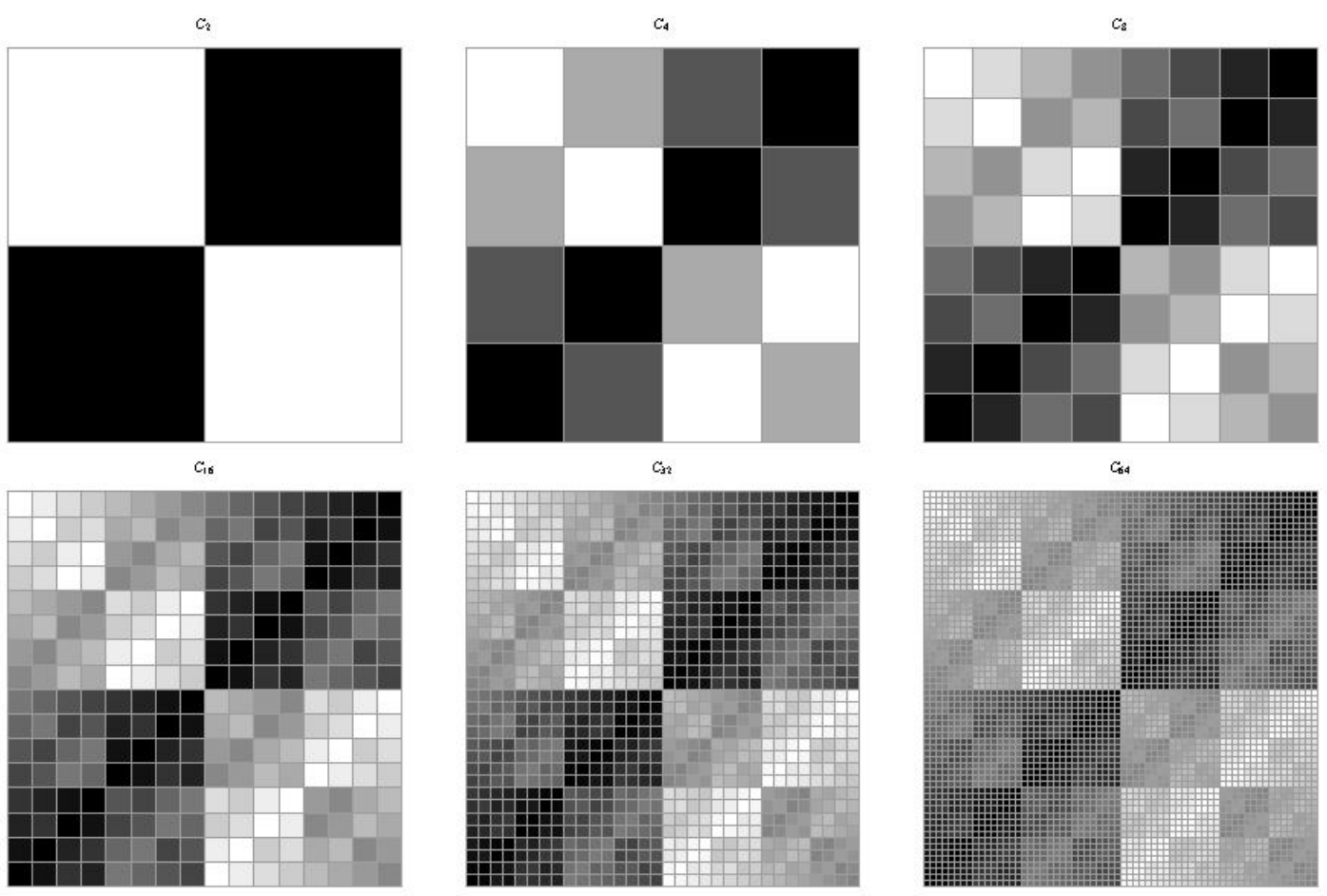

FIGURE 1. Group Multiplication Tables for Cyclic groups $C_{2^i}$, for $i = 1, \ldots, 6$.

for instance $O_3(1) = (1, 5, 3, 7, 0, 4, 2, 6)$

$$(O_3(x))_{0\leq x\leq 7} = \begin{pmatrix} 0 & 4 & 2 & 6 & 1 & 5 & 3 & 7 \\ 1 & 5 & 3 & 7 & 0 & 4 & 2 & 6 \\ 2 & 6 & 1 & 5 & 3 & 7 & 0 & 4 \\ 3 & 7 & 0 & 4 & 2 & 6 & 1 & 5 \\ 4 & 2 & 6 & 1 & 5 & 3 & 7 & 0 \\ 5 & 3 & 7 & 0 & 4 & 2 & 6 & 1 \\ 6 & 1 & 5 & 3 & 7 & 0 & 4 & 2 \\ 7 & 0 & 4 & 2 & 6 & 1 & 5 & 3 \end{pmatrix}.$$

Furthermore, we have also a formula defining the generator of $G_i$:

$$\mu_i = \bigotimes_{k=0}^{i} I_{2^k} \oplus 2^k. \tag{3.3}$$

where $\otimes$ denotes the concatenation and $\oplus$ denotes the componentwise sum.

*Example* 3.1. Let us compute the generators of the $G_i$'s:

$$\begin{aligned} &\mu_0 = (0) \\ &\mu_1 = ((0) \oplus 1) \otimes (0) = (10) \\ &\mu_2 = ((01) \oplus 2) \otimes (10) = (23) \otimes (10) = (2310) \\ &\mu_3 = ((0123) \oplus 4) \otimes (2310) = (4567) \otimes (2310) = (45672310) \end{aligned}$$

we note that for any $i$ we have

$$\mu_{i+1} = (I_{2^i} \oplus 2^i) \otimes \mu_i$$

and by iteratively expanding the above formula we get

$$\mu_{i+1} = (I_{2^i} \oplus 2^i) \otimes \mu_i = (I_{2^i} \oplus 2^i) \otimes (I_{2^{i-1}} \oplus 2^{i-1}) \otimes \mu_{i-1}.$$

From Theorem 2.4, we infer that

**Proposition 3.1.** *The von Neumann group algebra $\mathcal{N}(G)$, associated with the group $G$, is a MASA in $\mathcal{R}$.*

The fact that $\mathcal{N}(G)$ is a MASA constitutes another universal property of this model of computation. Indeed the construction we propose verifies two important properties:

- two any hyperfinite $\mathrm{II}_1$ factors are isomorphic;

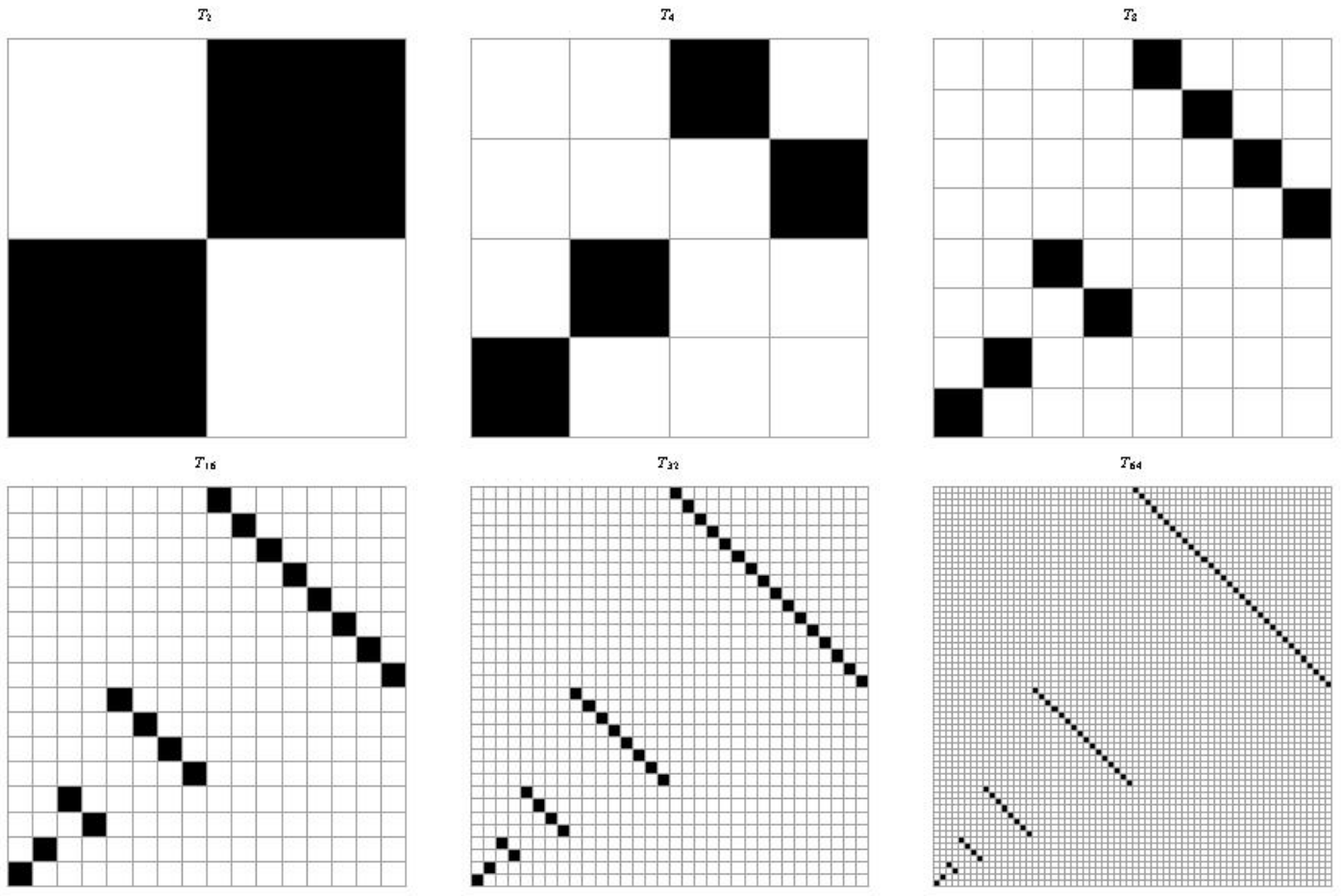

FIGURE 2. Matrix $T_i$, for $i = 1, \ldots, 6$.

- the abelian subalgebra we introduced is maximal.

Obviously it can be noted that the group $G$ can be replaced by $\mathbb{Z}$ as a denumerable discrete abelian group, since the corresponding von Neumann group algebras are isomorphic. Yet the very construction of the group $G$ plays a major role in the interpretation of the computational bound, for it gives an explicit *stratification* of subalgebras matching the stratification of the elementary complexity class. Although any another construction can do the same job and by isomorphism can be stratified in a similar manner, its description may turn out to be cumbersome.

## 4. Deterministic TMs encoded in a von Neumann Algebra

In this section, we study the encoding of TM's in operators acting on von Neumann group algebra $\mathcal{N}(G)$ of the group $G$, that we have introduced in the previous section. Leaving aside some obvious differences, our construction is similar to the one given in the work by Bernstein and Vazirani [BV97] and then reproposed by Nishimura and Ozawa [ON00, NO02].

There are many standard variations on the theme of deterministic Turing machines. Yet, for what concerns computable functions and the elementary complexity class, we may consider without lesser generality one-way infinite tape machines whose alphabet consists of a unique symbol. For further investigations on more general complexity questions (such as the case of polynomial computations), which are outside the scope of this work, we should have to consider an alphabet with not fewer than two symbols. Thus, for the sake of simplicity, here we deals with the unary case: $A = \{1\}$.

A one-way infinite tape TM with set of states $Q$ and alphabet $A$ is defined by the triplet $(A, Q, \mu)$, where $\mu$ is

$$\mu : Q \times A \to Q' \times \{-1, +1\} \times A',$$

with

$$A' = A \cup \{\Box\} = \{1, \Box\}.$$

We refer to the component of the triple $\mu(q, a) = (q', \epsilon, a)$ by subscript index: $\mu(q, a)_1 = q'$, $\mu(q, a)_2 = \epsilon$ and $\mu(q, a)_3 = a'$. The alphabet $A$ is extended to $A'$ by means of a special *blank* symbol. The set $Q$ of the states is extended to $Q'$ by adding a final state $q_F$ which marks the end of computation

$$Q' = Q \cup \{q_F\}.$$

Recall that there is a slight difference between Turing machines devised to represent computable functions (as in our case), where a unique final state suffices, and Turing machines associated with decision problems, where the decision of accepting or rejecting a certain input depends upon the final state reached between the two possible ones.

A *configuration*, or *instantaneous description* of the TM, is given by the state $q \in Q'$ of the finite control, the location of the tape head $p$, and a complete description of the contents of the tape. For a given alphabet $A$ and set of the states $Q$ we have the *space of configurations* which is

$$C(Q, A) = \{(q, p, f) \,|\, q \in Q', f : \mathbb{N} \to A', p \in \mathbb{N}\}.$$

As usual, at any time only a finite number of tape cells may contain nonblank symbols. For any TM $\mu$ and input tuple $x = (x_1, x_2, \ldots, x_n)$ of words $x_i \in A^*$ on its alphabet we determine the *computation of* $\mu$ *associated with* $x$ as a sequence (possibly infinite) of configurations $c_i \in C(Q, A)$:

$$S_\mu(x) := (c_0 = c_0(x), c_1, \ldots, c_n, \ldots)$$

of configurations which satisfy the following relations for any $i \in \mathbb{N}$:

$$c_{i+1} = \tau_\mu(c_i)$$

where $\tau_\mu : C(Q, A) \to C(Q, A)$ is the *transition function* which associates to any configuration $c_i = (q, p, f)$ its *successor configuration* $c_{i+1} = (q', f', p')$ where

$$q' = \mu(f(p), q)_1, \tag{4.1}$$

$$p' = p + \mu(f(p), q)_2 \tag{4.2}$$

$$f'(k) = \begin{cases} f(k) & \text{if } k \neq p, \\ \mu(f(p), q)_3 & \text{if } k = p. \end{cases} \tag{4.3}$$

By convention the initial configuration $c_0(x) = (q_0, 1, f)$ associated with the input $x = (x_1, x_2, \ldots, x_n)$ is such that the tape $f$ is:

(4.4)

$$f(k) = \begin{cases} x_i(k - \sum_{j=1}^{i-1} |x_j| + (1-i)) & \text{if } \sum_{j=1}^{i-1} |x_j| + (1-i) < k \leq \sum_{j=1}^{i} |x_j| + (1-i), \\ \Box & \text{otherwise.} \end{cases}$$

A *final* configuration is a configuration $(q_F, 1, f)$ for some tape $f$. In the theory of Quantum Turing Machine (QTM), the pair $(Q, A)$ is called *Turing frame* and the quantum state space of $(Q, A)$ is the Hilbert space spanned by the configuration space $C(Q, A)$ with the canonical basis $\{|c\rangle \; c \in C(Q, A)\}$ called the *computational basis*. Here, we add a further dimension to the interpretation by mapping the configuration space $C(Q, A)$ on the Hilbert spaces $\ell^2(G_i)$ underlying the group von Neumann algebras of the groups $G_i$ (that is, the groups which determine the locally finiteness of the group $G$).

This introduce a sort of *pre-quantum stratified Turing machine*, viewed in relation to the group von Neumann algebra $\mathcal{N}(G)$ which will be related to the Kalmar complexity in the next sections.

4.1. **Computational trees and branches.** Any computation for an $n$-ary elementary function $\varphi : \mathbb{N}^n \to \mathbb{N}$ on a given input $x = (x_1, \ldots, x_n)$ starts in an initial configuration $c_0(x) = (q_0, 1, f_0)$ and ends in a final configuration $c_F(y) = (q_F, 1, f_y)$, where $y = \varphi(x)$ and $f_y$ is the tape representing the integer $y$:

$$f_y(k) = \begin{cases} 1 & \text{if } 1 \leq k \leq y+1 \\ \square & \text{otherwise.} \end{cases}$$

We denote by $|(x_1, \ldots, x_n)|$ the *length of the input* $(x_1, \ldots, x_n)$ which is the length of the sequence encoding the input on the initial tape. It results that

$$|(x_1, \ldots, x_n)| = \sum_{i=1}^{n} x_i + (2n-1). \qquad (4.5)$$

This length is determinable because unary representation of integer encodes any integer number $x$ with a sequence of 1s of length $x+1$, and two consecutive integers in $x$ that, as usual, are separated by a blank on the tape.

*Proof.* The proof of (4.5) is by induction: if $n = 1$, then $|(x_1)| = x_1 + 1$ which is obtained by Equation (4.5) by taking $n = 1$, $\sum_{i=1}^{n} x_i + (2n-1) = \sum_{i=1}^{1} x_i + (2-1) = x_1 + 1$.

Then, it is easy to see that the length of $(x_1, \ldots, x_n, x_{n+1})$ is given by the length of the first $n$ elements to which we add 1 (because a blank separates $x_n$ and $x_{n+1}$) plus the length of $x_{n+1}$. Hence by inductive step we may conclude that:

$$|(x_1, \ldots, x_n, x_{n+1})| = |(x_1, \ldots, x_n)| + 1 + |(x_{n+1})| = \sum_{i=1}^{n} x_i + (2n-1) + 1 + (x_{n+1} + 1) =$$

$$= \sum_{i=1}^{n+1} x_i + (2n-1) + 1 + 1 = \sum_{i=1}^{n+1} x_i + (2(n+1) - 1).$$

□

For any final value $y$, let us denote by

$$\mathsf{inputs}_L(y) := \{x \text{ s.t. } |x| = L \text{ and } S_\mu(x) \text{ ends with } c_F(y)\}$$

the set of input vectors leading to the same final configuration.

By taking into account the set of computations with a fixed output $y$ and input of length at most $L$, we define the *computational tree* as follows:

$$T_L(y) = \{S_\mu(x) |\ \varphi(x) = y \text{ and } |x| = L\}.$$

In particular, a computational tree is *regular*, if all its computational branches are of the same length. At any rate, it is important to note that in the computational tree $T_L(y)$ associated with $y$, any computational branch corresponds to a computation starting in one particular input $x$ (of length $L$), for which $\varphi(x) = y$.

For any computational tree $T_L(y)$, the associated computational *virtual tree* $VT_L(y)$ is defined as a regular tree with the same number of computational branches as $T_L(y)$ and where all the branches have the same length $h = max_{x \in \mathsf{inputs}_L(y)} |S_\mu(x)|$. Short branches in $T_L(y)$ are completed by virtual configurations $c_0^{-k}(x)$ with $k = 1, \ldots, k_x$ where $k_x = h - |S_\mu(x)|$. We may put $c_0^0(x) = c_0(x)$ for the sake of a uniform notation.

Furthermore let us denote by

$$\mathsf{leaves}(VT_L(y)) := \{c_0^{-k_x}(x) |\ x \in \mathsf{inputs}_L(y)\ \}$$

the set of the (virtual) initial configurations associated with the input vectors in $\mathsf{inputs}_L(y)$.

Moreover, the symbol $B_L(x)$ stands for the computational branch associated with the input $x$ in the virtual tree $VT_L(y)$. It follows that a computational branch is a sequence of configurations or virtual configurations, taken in:

$$VC(Q,A) := C(Q,A) \cup \{c_0^{-k}(x) |\ x \in \mathsf{inputs}_L(y) \text{ and } 1 \leq k \leq k_x\}.$$

We consider also subtrees of $VT_L(y)$ rooted at any $c \in VT_L(y)$ and we denote by $\mathsf{inputs}_c(VT_L(y))$ the subset of those elements in $\mathsf{inputs}_L(y)$ whose computational branch intersects the subtree rooted in $c$.

For any virtual tree $VT_L(y)$ we define a mapping from the disjoint union of its computational branches and the elements of the basis of a Hilbert space. Since any computational branch is associated with an input, we can denote an element of this disjoint union by a pair in $VC(Q,A) \times \mathsf{inputs}_L(y)$.

$$[\,] : VC(Q,A) \times \mathsf{inputs}_L(y) \to \mathcal{H}.$$

We end this subsection by introducing two further notions: the function $\mathsf{depth}_{VT_L(y)}(c)$ of a configuration $c$ in $VT_L(y)$, which is simply defined as the distance from the final configuration $c_F(y)$; the $\mathsf{level}_{VT_L(y)}(c)$, that is the set of configurations at the same depth of $c$:

$$\mathsf{level}_{VT_L(y)}(c) := \{c' \in VT_L(y) \text{ such that } \mathsf{depth}_{VT_L(y)}(c') = \mathsf{depth}_{VT_L(y)}(c)\}.$$

Finally, for any input $x \in \mathsf{inputs}_L(y)$, we define a map from configurations in $B_L(x) \subset VT_L(y)$ to elements of the Hilbert space $\mathcal{H}$:

$$[[c]]_x := \sum_{c' \in \mathsf{level}_{VT_L(y)}(c)} \ \sum_{x' \in \mathsf{inputs}_{c'}(VT_L(y))} \alpha(c',x')[(c',x')] \tag{4.6}$$

where

$$\alpha_x(c',x') := \frac{\sqrt{1+\delta(x,x')}}{\sqrt{|\mathsf{leaves}(VT_L(y))|+1}}$$

and $\delta$ is the Kronecker delta function

$$\delta(x,x') := \begin{cases} 1 & \text{if } x = x', \\ 0 & \text{if } x \neq x'. \end{cases}$$

4.2. **Digression on counting and bounding initial configurations.** In this subsection we seek an estimation of the number of the possible initial configurations for a Turing machine corresponding to an $n$-ary input vector of integers with a unary representation of length $L$.

The solution to this problem is a matter of simple combinatorial computation: since by definition any initial configuration is $(q_0, 1, f)$, the parameter is given by the initial tape itself . Thus, the question now becomes how to compute the number of sequences $w$ of length $L$, $w \in \{\Box, 1\}^L$, such that the following conditions hold:

(1) $w$ contains exactly $n-1$ blanks;
(2) any $\Box$ in $w$ cannot follow another $\Box$;
(3) $w$ cannot start or end with $\Box$.

It turns out that this number of sequences can be determined by the following recursive formula (related to the Fibonacci sequence):

$$f(1,1) = 1 \tag{4.7}$$

$$f(2,1) = 1 \tag{4.8}$$

$$f(L+1,n) = f(L,n) + f(L-1,n-1) \tag{4.9}$$

*Proof.* (by induction). Evidently there is only one sequence of length $L = 1$ without any blank symbol (i.e., representing an input vector of length 1). If we admit an input of length $L = 2$, we may again consider only vectors of length 1, because otherwise we would get an initial or final blank; thus, there is only one sequence formed by two 1's and no blanks.

Let us to come to the induction step. If we extend the input to the length $L + 1$, we get all the sequences obtained with length $L$ extended by an extra 1, which are $f(L, n)$. On the other hand, we may also put a blank in position $L$ because the extra 1 placed at the end of the word makes these words admissible. Since we have used one blank and the last two positions are fixed, we count all the possible inputs of length $L - 1$ representing a vector of length $n - 1$, and consequently $f(L - 1, n - 1)$. □

Note that the following holds

$$Fib(L) = \sum_{n=1}^{\lceil L/2\rceil+1} f(L,n). \tag{4.10}$$

*Proof.* This proof is also by induction on $L$.

$$Fib(1) = \sum_{n=1}^{\lceil 1/2\rceil+1} f(1,n) = f(1,1) + f(1,2) = 1,$$

$$Fib(2) = \sum_{n=1}^{\lceil 2/2\rceil+1} f(2,n) = f(2,1) + f(2,2) = 1.$$

$$Fib(L+1) = Fib(L) + Fib(L-1) = \sum_{n=1}^{\lceil L/2\rceil+1} f(L,n) + \sum_{n=1}^{\lceil (L-1)/2\rceil+1} f(L-1,n) =$$
$$\sum_{n=1}^{\lceil L/2\rceil+1} f(L,n) + \sum_{n=1}^{\lceil (L-1)/2\rceil+1} f(L-1,n-1)$$

We now distinguish the two cases of even or odd $L$.

If $L = 2k$, then $\lceil L/2\rceil + 1 = k + 1$ and also $\lceil (L-1)/2\rceil + 1 = k + 1$ by effect of the approximation. Hence the two sums have the same number of terms, namely $k + 1$, and so we obtain:

$$\sum_{n=1}^{\lceil L/2\rceil+1} f(L,n) + f(L-1,n-1) = \sum_{n=1}^{\lceil L/2\rceil+1} f(L+1,n).$$

Note that this equality not yet yields 4.10 for the bound of the sum should be $\lceil (L+1)/2\rceil + 1 = \lceil (2k+1)/2\rceil + 1 = k + 2$, and so a term in the sum is missing. Anyway, the missing term is $f(L+1, k+2) = f(L+1, \lceil (L+1)/2\rceil + 1)$ which we can prove to be zero for any $L$

$$\sum_{n=1}^{\lceil L/2\rceil+1} f(L+1,n) = \sum_{n=1}^{\lceil L/2\rceil+1} f(L+1,n) + f(L+1,\lceil (L+1)/2\rceil+1) = \sum_{n=1}^{\lceil (L+1)/2\rceil+1} f(L+1,n).$$

If $L = 2k+1$, then $\lceil L/2\rceil+1 = k+2$ by effect of the approximation and $\lceil (L-1)/2\rceil+1 = k+1$. Thus, the two sums do not have the same number of terms, namely $k+2$ and $k+1$. Anyway, the missing term in the second sum is $f(L+1,k+2) = f(L+1,\lceil (L+1)/2\rceil+1)$ which again is zero for any $L$. Hence, we obtain:

$$\sum_{n=1}^{\lceil L/2\rceil+1} f(L,n) + \sum_{n=1}^{\lceil (L-1)/2\rceil+1} f(L-1,n-1) + f(L+1,\lceil (L+1)/2\rceil+1) = \sum_{n=1}^{\lceil L/2\rceil+1} f(L+1,n).$$

□

Consequently, a simple bound for the number of initial configurations of length $L$ for a TM representing a $n$-ary function is

$$f(L,n) \leq Fib(L).$$

Our knowledge of the bound can be improved by making explicit that $f(L,n)$ is the following binomial coefficient:

$$f(L,n) = \binom{L-n+1}{n-1} = \frac{(L-n+1)!}{(L-n+1-(n-1))!(n-1)!}.$$

Then, by considering that for any $L$ the maximal value of $f(L,n)$ is given by $n$ yielding the two inequalities, $f(L,n-1) > f(L,n)$ and $f(L,n) < f(L,n+1)$. Both conditions depend on the ratio $\dfrac{f(L,n-1)}{f(L,n)}$ that can be simplified insofar as it is the ratio between two binomial coefficients:

$$\frac{f(L,n-1)}{f(L,n)} = (n+1)\frac{(L-n)}{(L-2n-1)(L-2n)}.$$

The searched value of $n$ is the solution of the inequality

$$(n+1)\frac{(L-n)}{(L-2n-1)(L-2n)} > 1$$

which is obtained when

$$n > \frac{1}{10}(7 + 5L - \sqrt{9 + 10L + 5L^2}).$$

Finally, we can establish the asymptotic behaviour of

$$\frac{1}{10}(7 + 5L - \sqrt{9 + 10L + 5L^2})$$

by computing its asymptote as

$$\frac{1}{10}(7 - \sqrt{5} + (5 - \sqrt{5})L) = \frac{1}{10}(7 - \sqrt{5}) + \frac{L}{A+2} \simeq 0.476393 + \frac{L}{A+2}$$

where $A = \dfrac{1+\sqrt{5}}{2}$.

The following table gives the values of $f(L, n)$ for the first $L$:

| | | | | | | | | | | |
|---|---|---|---|---|---|---|---|---|---|---|
| **1** | | | | | | | | | | |
| **1** | 0 | | | | | | | | | |
| **1** | 1 | | | | | | | | | |
| 1 | **2** | 0 | | | | | | | | |
| 1 | **3** | 1 | | | | | | | | |
| 1 | **4** | 3 | 0 | | | | | | | |
| 1 | 5 | **6** | 1 | | | | | | | |
| 1 | 6 | **10** | 4 | 0 | | | | | | |
| 1 | 7 | **15** | 10 | 1 | | | | | | |
| 1 | 8 | **21** | 20 | 5 | 0 | | | | | |
| 1 | 9 | 28 | **35** | 15 | 1 | | | | | |
| 1 | 10 | 36 | **56** | 35 | 6 | 0 | | | | |
| 1 | 11 | 45 | **84** | 70 | 21 | 1 | | | | |
| 1 | 12 | 55 | 120 | **126** | 56 | 7 | 0 | | | |
| 1 | 13 | 66 | 165 | **210** | 126 | 28 | 1 | | | |
| 1 | 14 | 78 | 220 | **330** | 252 | 84 | 8 | 0 | | |
| 1 | 15 | 91 | 286 | **495** | 462 | 210 | 36 | 1 | | |
| 1 | 16 | 105 | 364 | 715 | **792** | 462 | 120 | 9 | 0 | |
| 1 | 17 | 120 | 455 | 1001 | **1287** | 924 | 330 | 45 | 1 | |
| 1 | 18 | 136 | 560 | 1365 | **2002** | 1716 | 792 | 165 | 10 | 0 |

By numerical computation we can give the following inequality:

$$f(L, -L\sqrt{5} + 4.76392) \leq 2^{\lambda(L)} \tag{4.11}$$

where $\lambda(L) = \dfrac{L}{0.890228A} - A$.

At this point we are in position to take into account the dimension $2^i$ of the group $G_i$ required in order to injectively embed any computational branch of length bounded by $N$.

Let us consider an input of length $L$. From equality (4.11), we infer that there are at most $2^{\lambda(L)}$ distinct of such inputs. For every such input the length of the associated computational branch is upper bounded by $h$ (namely, $h$ is the depth of $VT_L(y)$) thus proving that the group $G_N$ has the suitable dimension $2^N$ whenever

$$N > \log(h) + \lambda(L). \tag{4.12}$$

## 5. Configurations in the Hilbert space $\ell^2(G)$

Let us fix the computational basis in the Hilbert space $\mathcal{H}$ within Equation 4.6. We take $\mathcal{H} = \ell^2(G)$ and we associate any pair $(c, x) \in VT_L(y) \times \mathsf{inputs}_L(y)$ with an element $g$ of $G$ in such a way that whenever $c' = \tau_\mu(c)$, i.e., the successor configuration, the corresponding element in $G = \bigoplus_{i=0}^{\infty} G_i$ is

$$[(c', x)] := \mu_N(g) \text{ where } (c', x) \in VT_L(y) \times \mathsf{inputs}_L(y)$$

obtained by the action of the $N$-th group generator $\mu_N$ on $g$. To safeguard injectivity we focus on the dimension $N$ of the group $G_N$ as discussed in the previous section, inequality (4.12), i.e., such that $N > \log(h) + \lambda(L)$. Note that the initial configurations have to be mapped on elements which are "far enough" (i.e., at least at distance $h$) in the orbit described by the generator $\mu_N$.

For any element in $\sum_{g\in G}\alpha_g g \in \ell^2(G)$, we define the set of the elements of $G$ with non-zero coefficients:

$$\mathsf{supp}(\sum_{g\in G}\alpha_g g) := \{g' \in G | \alpha_{g'} \neq 0\}.$$

For any input $x \in \mathbb{N}^n$, we consider the one-to-one mapping $[[_]]_x : VT_L(\varphi(x)) \to \ell^2(G)$ which sends each $c \in VT_L(\varphi(x))$ to the element $[[c]]_x$ as in Definition 4.6.

**Definition 6.** $[[x]]$ denotes the interpretation in $\ell^2(G)$ of the initial configuration associated with $x$:

$$[[x]] := \mu_N^{k_x}([[c_0^{-k_x}(x)]]_x).$$

It is worth noting that the element $[[x]]$ is obtained by considering the leaf $c_0^{-k_x}(x)$ of the virtual tree $VT_L(\varphi(x))$ on the same computational branch $B_L(x)$ of the initial configuration $c_0(x)$ associated with $x$. Moreover, by iterating $k_x$ times the application of $\mu_N$ to the element $[[c_0^{-k_x}(x)]]_x \in \ell^2(G)$ comes to coincide with a combination of elements of $\ell^2(G)$ associated with configurations at the same level of $c_0(x)$ in the virtual tree. Furthermore, the interpretation $[[x]]$ exploits the fact that the action of $\mu_N$ (the generator of the group $G_N$) on $[[x]]$ iterated $h_x$ time (where $h_x$ is the depth of the computational branch $B_L(x)$ in $T_L(\varphi(x))$) leads to the final configuration $[c_F(\varphi(x))]_x$.

**Proposition 5.1.** *For any pair $x, x' \in \mathsf{inputs}(VT_L(y))$ and such that $h_x = h_{x'}$,*

*a)* $\mathsf{supp}([[x]]) = \mathsf{supp}([[x']])$*;*
*b)* $\mathsf{supp}(\mu_N^k([[x]])) = \mathsf{supp}(\mu_N^k([[x']]))$ *for any $k \leq h_x$.*

*Proof of a).* By definition

$$[[x]] = \mu_N^{k_x}([[c_0^{-k_x}(x)]]_x),$$

and analogously

$$[[x']] = \mu_N^{k_{x'}}([[c_0^{-k_{x'}}(x')]]_{x'}).$$

Under the hypothesis that $h_x = h_{x'}$, it holds that $k_x = h - h_x = h - h_{x'} = k_{x'}$. Therefore, the configurations $\mathsf{level}_{VT_L(y)}(c_0^{-k_x}(x))$ at the same depth of $c_0^{-k_x}(x)$ and the ones $\mathsf{level}_{VT_L(y)}(c_0^{-k_{x'}}(x'))$ at the same depth of $c_0^{-k_{x'}}(x')$ coincide. Let us denote such a set of nodes by $A$.

By Definition 4.6, we have

$$[[c_0^{-k_x}(x)]]_x = \sum_{c'\in A}\sum_{w\in \mathsf{inputs}_{c'}(VT_L(y))} \alpha_x(c', w)[(c', w)]$$

and

$$[[c_0^{-k_{x'}}(x')]]_{x'} = \sum_{c'\in A}\sum_{w\in \mathsf{inputs}_{c'}(VT_L(y))} \alpha_{x'}(c', w)[(c', w)].$$

Thus, in the two sums we have a combination of the same terms having non null coefficients although possibly different $\alpha_x(c', w)$ and $\alpha_{x'}(c', w)$. This implies that also the two supports coincide. □

*Proof of b).* By linearity of $\mu_N$ applied to

$$\begin{aligned}
\mu_N^k([[x]]) &= \mu_N^k(\mu_N^{k_x}([[c_0^{-k_x}(x)]]_x)) = \\
&= \mu_N^{k+k_x}([[c_0^{-k_x}(x)]]_x) = \\
&= \mu_N^{k+k_x}(\sum_{c'\in A}\sum_{w\in \mathsf{inputs}_{c'}(VT_L(y))} \alpha_x(c',w)[(c',w)] = \\
&= \sum_{c'\in A}\sum_{w\in \mathsf{inputs}_{c'}(VT_L(y))} \alpha_x(c',w)\mu_N^{k-k_x}([[(c',w)])
\end{aligned}$$

and in a similar way for $x'$ we have

$$\mu_N^k([[x']]) = \sum_{c'\in A}\sum_{w\in \mathsf{inputs}_{c'}(VT_L(y))} \alpha_{x'}(c',w)\mu_N^{k-k_{x'}}([[(c',w)])$$

and being $k_x = k_{x'}$, again the two supports come to coincide. □

**Proposition 5.2.** *For any* $x, x' \in \mathsf{inputs}(VT_L(\varphi(x))$*,*

*a)* $||(\mu_N^k([[x]]))|| = 1$*,*
*b)* $\langle \mu_N^k([[x]]), [(c,x)]\rangle = \sqrt{2}\langle \mu_N^{k+(h_x-h_{x'})}([[x']]), [(c,x)]\rangle$ *if* $c \in S_\mu(x)$*,*

*for any* $k \leq h_x$*.*

*Proof of a).* Let us recall the definition of the norm: since we consider elements of $\ell^2(G)$, we compute the norm of $h = \sum_{g\in G}\alpha_g g$ as $||h|| = \left(\sum_{g\in G}\alpha_g^2\right)^{1/2}$.

Again by Definition 4.6, we have

$$[[c_0^{-k_x}(x)]]_x = \sum_{c'\in A}\sum_{w\in \mathsf{inputs}_{c'}(VT_L(y))} \alpha_x(c',w)[(c',w)]$$

When we compute the square of any element in the sum, by definition of $\alpha_x(c',w)$ we have that $\alpha_x(c',w)^2 = \dfrac{1+\delta(x,x')}{|\mathsf{leaves}(VT_L(y))|+1}$. Moreover, since the sum is indexed on the set

$$\mathsf{leaves}(VT_L(y)) = \bigcup_{c'\in \mathsf{level}_{VT_L(y)}(c)} \mathsf{inputs}_{c'}(VT_L(y))$$

it follows that $\delta(x,x') = 1$ only whenever $x' = x$ so that the numerator in $\alpha_x(c',w)$ is always 1 except for one term where its value is $\sqrt{2}$.

As a consequence of the fact that the leaves of the computational tree bijectively correspond to inputs of the computational tree, one has that

$$\sum_{c' \in \mathsf{level}_{VT_L(y)}(c)} \sum_{x' \in \mathsf{inputs}_{c'}(VT_L(y))} \alpha_x(c', x')^2 =$$

$$= \sum_{c' \in \mathsf{level}_{VT_L(y)}(c)} \sum_{x' \in \mathsf{inputs}_{c'}(VT_L(y))} \frac{1 + \delta(x, x')}{|\mathsf{leaves}(VT_L(y))| + 1} =$$

$$= \frac{1}{|\mathsf{leaves}(VT_L(y))| + 1} \left( \sum_{\substack{x' \in \mathsf{inputs}(VT_L(y)) \\ x' \neq x}} 1^2 + \sqrt{2}^2 \right) =$$

$$= \frac{1}{|\mathsf{leaves}(VT_L(y))| + 1} (|\mathsf{inputs}(VT_L(y)) \backslash \{x\}| + 2) =$$

$$= \frac{|\mathsf{inputs}(VT_L(y))| + 1}{|\mathsf{leaves}(VT_L(y))| + 1} = 1.$$

f To end the proof of the claim it is enough to recall that $[[x]] = \mu_N^{k_x}([[c_0^{-k_x}(x)]]_x)$ and the linearity of the action of $\mu_N$, which acts isometrically on $[[c_0^{-k_x}(x)]]_x$, thus,

$$||[[x]]|| = ||[[c_0^{-k_x}(x)]]_x|| = 1.$$

For the same reason we also get part $a$):

$$||\mu_N^k([[x]])|| = ||[[x]]|| = 1.$$

□

*Proof of b).* Part $b$) follows directly by observing that the scalar product takes as a value the coefficient of the term $[(c, x')]$ when it belongs to the support of $[[x]]$. This may happen if the configuration $c \in \mathsf{level}_{VT_L(y)}(c_0(x))$ and $x' \in \mathsf{inputs}_{VT_L(y)}(c)$. Under these conditions one has by Definition 4.6 that

$$\langle [[x']], [(c, x)] \rangle = \alpha_{x'}(c, x) = \frac{\sqrt{1 + \delta(x, x')}}{\sqrt{|\mathsf{leaves}(VT_L(y))| + 1}}.$$

Thus, if $x \neq x'$ we have

$$\langle [[x']], [(c, x)] \rangle = \alpha_{x'}(c, x) = \frac{1}{\sqrt{|\mathsf{leaves}(VT_L(y))| + 1}}$$

and

$$\langle [[x]], [(c, x)] \rangle = \alpha_x(c, x) = \frac{\sqrt{2}}{\sqrt{|\mathsf{leaves}(VT_L(y))| + 1}}$$

so that we the thesis follows in the case $k = 0$ and $h_x = h_{x'}$.

If $k \neq 0$ (or respectively $h_x \neq h_{x'}$), then, by construction (or resp. by definition) we have to apply $\mu_N$ to $[[x]]$ and $[[x']]$; as in the previous case, the coefficient of $[(c, x)]$ coincides with the coefficient of $\mu_N([(c, x)])$ and by linearity in the sum we conclude that

$$\alpha_x(\mu_N([(c, x)])) = \sqrt{2}\alpha_{x'}(\mu_N([(c, x)])).$$

□

By definition, it is easily seen that whenever a computation $S_\mu(x)$ of the TM $\mu$ associated with an input $x$ reaches a configuration $c_i$ and the successor configuration $\mu(c_i) = c_{i+1}$, we have that

$$\mu_N([[c_i]]_x) = [[c_{i+1}]]_x. \tag{5.1}$$

Thus it also holds true that $\mu_N(\mathsf{supp}([[c_i]]_x)) = \mathsf{supp}([[c_{i+1}]]_x)$. This implies that for a different input $x'$ which contains $c_i$ in the computation $S_\mu(x')$ we obtain two different interpretations associated with the same $c_i$, i.e., $[[c_i]]_{x'} \neq [[c_i]]_x$. Strictly speaking this fact tells us that such model yields a non-uniform interpretation. On the other side, by Proposition 5.1, it follows that

$$\mathsf{supp}([[c_i]]_{x'}) = \mathsf{supp}([[c_i]]_x). \tag{5.2}$$

This amounts to saying that in view of a uniform interpretation, we cannot exploit $\mathsf{supp}([[c]]_x)$ for the interpretation of the configuration $c$ since the support of two different inputs $x \neq x'$ comes to coincide under different interpretations (against injectivity). Yet if we restrict our attention to the support for the *final* configurations we get a correct interpretation (injectivity holds). As concerns these final configurations, indeed, we can consider their interpretation modulo the application of $\mathsf{supp}$ because it is impossible that $c_F(\varphi(x)) = c_F(\varphi(x'))$, if $\varphi(x) \neq \varphi(x')$.

The following diagram recapitulates this fact:

$$\begin{array}{ccc} C(A,Q) & \xrightarrow{\tau_\mu^h} & C(A,Q) \\ \big\downarrow [[.]]_x & & \big\downarrow \mathsf{supp}([[.]]_x) \\ \ell^2(G) & \xrightarrow{\mu_N^h} & \ell^2(G). \end{array}$$

Such commutativity enables us to prove the following proposition, where *configurations* of the Turing machine correspond to *elements* of the canonical basis of the Hilbert space $\ell^2(S_X)$ underlying the hyperfinite $\mathrm{II}_1$ factor $\mathcal{R} \simeq \mathcal{N}(S_X)$ (see Proposition 2.3):

**Proposition 5.3.** *Let $\varphi \in \mathrm{DSPACE}(S(n))$ and let $\mu$ be a Turing machine computing $\varphi$. Then, for each input $x$, $||x|| \leq n$ there is an interpretation $[\mu] \in \mathcal{R}$ with a mapping $[[.]]_x : C(A,Q) \to \ell^2(S_X)$ such that*

*a) $[\mu]([[c_j]]_x) = [[\tau_\mu(c_j)]]_x$ where $\tau_\mu$ is the transition function associated with $\mu$;*
*b) $\mathsf{supp}([\mu]^k([[x]])) = \mathsf{supp}([c_F(\varphi(x))]_x)$ for $k = |S_\mu(x)|$.*

*Proof.* Let us consider the interpretation $[[.]]_x : C(A,Q) \to \ell^2(G)$ in Definition 6. We define $[\mu] := J(\mu_{S(N)})$ the image of the operator $\mu_{S(N)}$ associated with the generator of the group $G_{S(N)}$ by the embedding $J : \mathcal{N}(G) \to \mathcal{R}$ determined by Theorem 2.4. Then, since $\mu_{S(N)}$ satisfies Equation 5.1), by isomorphism $[\mu] = J(\mu_{S(N)})$ verifies the same property as well. Thus part $a)$ is proven.

As concerns part $b)$, again by Equation 5.1 it easily checked that $[\mu]([[c]]) = [[\tau_\mu(c)]]$. If we iterate the application of $[\mu]$, we get $[\mu]^k([[c]]) = [[\tau_\mu^k(c)]]$ and the equality also holds if we pass to the support:

$$\mathsf{supp}([\mu]([[c]])) = \mathsf{supp}([[\tau_\mu(c)]]).$$

□

5.1. **Space Bounded TMs.** Now we denote by $\mathcal{E}$ the class of elementary functions, defined and investigated by Kalmár [Kal43]. $\mathcal{E}$ is the least class of primitive recursive functions that contains the constant 0, all projections, successor, addition, cut-off subtraction and multiplication, and closed under composition, bounded sum and bounded product.

Since $\mathcal{E}$ is closed under composition, for each $m$ the $m$-times iterated exponential $2^{[m]}(x)$ is in $\mathcal{E}$, where $2^{[m+1]}(x) = 2^{2^{[m]}(x)}$ and $2^{[0]}(x) = x$. The elementary functions are exactly the functions computable in elementary time, i.e., the class of functions computable by a TM in a number of steps bounded by some elementary function. Two results are well known concerning the class $\mathcal{E}$:

**Proposition 5.4.** *(1)* $\mathcal{E} = \mathrm{DTIME}(\mathcal{E}) = \mathrm{DSPACE}(\mathcal{E})$.

*(2) If $\varphi \in \mathcal{E}$, then there is a number $m$ such that for all $x = (x_1, \ldots, x_n)$,*

$$\phi(x) \leq 2^{[m]}(|x|).$$

However, Proposition 5.4.2 tell us that $\mathcal{E}$ does not contain the iterated exponential $2^{[m]}(|x|)$ where the number of iterations $m$ is a variable, since any function in $\mathcal{E}$ has an upper bound where $m$ is *fixed*. This remark is very useful to obtain the following result:

**Proposition 5.5.** *Let $\mu$ be a TM computing a Kalmar elementary function $\varphi \in \mathcal{E}$. Then, there exists an integer $m$ such that for every input $x$, $|x| \leq n$, the computation of $\mu$ starting from the initial configuration associated with $x$ is representable in $\mathcal{N}(G_M)$ with $M = 2^{[m]}(n)$ for some $m$ that does not depend upon $n$.*

*Proof.* From Proposition 5.4.1, we get that since $\mu$ computes $\varphi \in \mathcal{E}$ there exists $\psi \in \mathcal{E}$ such that $\varphi \in \mathrm{DSPACE}(\psi(n))$. As a result, for every input $x$, $|x| \leq n$, the machine $\mu$ has halt space $s \leq \psi(n)$. By Proposition 5.4.2, this implies that there exists $m^*$ such that $s \leq 2^{[m^*]}(|x|)$ where $m*$ does not depend upon $|x|$.

For Proposition 5.3, by choosing $m := m^* - 1$, we have that the computation of the machine $\mu$ starting from the initial configuration associated with $x$ is representable in $\mathcal{N}(G_M)$ where $M = 2^{[m]}(n)$, thus also $m$ does not depend upon $n$. □

## Acknowledgements

We would like to thank Sebastiano Carpi for various stimulating discussions on some topics of this paper.

Marco Pedicini, Consiglio Nazionale delle Ricerche, Istituto per le Applicazioni del Calcolo Mauro Picone, Viale Manzoni 30, 00185 Roma, Italy

*E-mail address*: marco@iac.cnr.it

Mario Piazza, Dipartimento di Filosofia, Università di Chieti-Pescara, Via dei Vestini 31, 66013 Italy

*E-mail address*: m.piazza@unich.it